\documentclass[showpacs,showkeys,twoside,eqsecnum,prd,twocolumn]{revtex4-1}
\usepackage{amsfonts,amsmath,amssymb,bm,hyperref,graphicx}

\begin{document}


\title{Spin-down of neutron stars by neutrino emission}

\author{Maxim Dvornikov${}^{a,b}$}
\email{maxim.dvornikov@usm.cl}
\author{Claudio Dib${}^{a}$}
\email{claudio.dib@usm.cl}
\affiliation{${}^{a}$Centro Cient\'{\i}fico-Tecnol\'ogico de
Valpara\'{\i}so and Departamento de F\'{i}sica, Universidad
T\'{e}cnica Federico Santa Mar\'{i}a,
Casilla 110-V, Valpara\'{i}so, Chile;\\
${}^{b}$IZMIRAN, 142190, Troitsk, Moscow Region, Russia}

\date{\today}

\begin{abstract}
We study the spin-down of a neutron star during its early stages
due to the neutrino emission. The mechanism we consider is the
subsequent collisions of the produced neutrinos with the outer
shells of the star. We find that this mechanism can indeed slow
down the star rotation but only in the first tens of seconds of
the core formation, which is when the appropriate conditions of
flux and collision rate are met. We find that this mechanism can
extract less than 1\% of the star angular momentum, a result which
is much less than previously estimated by other authors.
\end{abstract}

\pacs{97.60.Bw, 14.60.Lm, 26.60.Gj}

\keywords{supernova neutrinos; rotating neutron stars}

\maketitle

\section{Introduction}

Neutrinos play a significant role in the evolution of various
astronomical objects. They should carry away almost all of the
gravitational energy lost in the collapse of a massive star during
a supernova explosion~\cite{GiuKim07p511}, phenomenon that was
confirmed by the detection of neutrinos from supernova SN1987A, in
the available experiments at the time, namely
Kamiokande-II~\cite{Hir87}, IMB~\cite{Bio87}, Baksan~\cite{Ale87},
and LSD~\cite{Dad87}. Neutrinos are also important in the cooling
of the neutron star formed at the center of a supernova
explosion~\cite{Lattimer91,YakPet04}. Neutrinos, because of their
weak interaction with matter, provide important signals of the
inner parts of stars like our Sun~\cite{Bahcall} and of the cores
in stellar collapses~\cite{Liebendorfer}.  Asymmetric neutrino
emission may also be responsible for the large peculiar velocities
observed in pulsars~\cite{Kusenko,Lai04}. It has also been
proposed a long time ago that neutrino emission can slow down the
spinning of neutron stars~\cite{Mik77,Eps78a}. Here we want to
revisit the latter idea, trying to refine the estimate by using
our current knowledge of the physics of neutrinos and their
emission in neutron stars.

The majority of neutron stars are known to have large angular
velocities, and in the case of radio pulsars one can directly
measure their speed of rotation. It is also observed that, on
average, their rotation tends to slow down with time, a phenomenon
that is explained by emission of electromagnetic waves or, in some
conditions, by the emission of gravitational waves or other
processes~\cite{spindown}. This should be the case during most of
the life of the neutron star. However, at the early stages, during
the collapse and formation of the star core (in a time scale of
$10\thinspace\mathrm{s}$), which is when the flux is intense
enough and the mean free path is comparable to the star
size~\cite{Raf96}, the spin-down of this protoneutron star (PNS)
can be influenced by the collisions of the neutrinos with the star
matter as they escape. It is interesting to notice that the
opposite effect, i.e. the acceleration of a neutron star rotation,
has also been proposed~\cite{GvoOgn02} for the case of neutrinos
interacting with strong toroidal magnetic fields inside stars.

Note that neutrino emission can also decrease the angular momentum
of the star due to the change in the gravitational mass of the
star~\cite{Eps78b,BauSha98,Jan04}. According to the estimate of
Ref.~\cite{Jan04}, a star can lose at most $40\%$ of its initial
angular momentum. Other proposed mechanisms of angular momentum
loss at early stages of the PNS evolution include viscous
processes~\cite{Tho05}, transfer of rotational energy into the
energy of the supernova explosion~\cite{AkiWhe05}, magnetic PNS
winds~\cite{Tho04}, and propeller mechanisms~\cite{Heg05}.
However, it was found in Ref.~\cite{Ott06} that none of these
theoretical explanations can robustly spin-down a PNS from about
several ms to the observed periods of rotation of young pulsars.
We must add that the loss of angular momentum by neutrino emission
may also occur if there is anisotropy at the neutrino production
points, which may be the case if the star matter has significant
polarization due to the star magnetic field~\cite{Dor84}. However,
we do not consider this effect here. In our work we are interested
in the case in which the production is rather isotropic, but
neutrinos travel a sizable distance in the star and subsequently
scatter with matter in outer shells where the transverse velocity
of the medium is larger.

In this work we revise the previous estimates for the spin-down of
PNS by neutrino emission~\cite{Mik77,Eps78a}, where it was stated
that this mechanism can even possibly stop the star rotation. This
result denotes a very dramatic effect, which we want to study
using a more detailed work, but still within analytical models in
order to study the sensitivity of our results to different
parameters of a generic star. Our estimates show that the effect
is much weaker than previous estimates -- less than 1\% reduction
of the angular speed. We also check that most refinements in our
treatment point to further reduction, not enhancement, of the
effect.

In Sec.~\ref{AMT} we present our conceptual treatment and
calculation of the spin-down of PNS due to neutrino emission. In
Sec.~\ref{CONCL} we summarize our results and state our
conclusions.


\section{Model and Calculations\label{AMT}}

In this section we formulate the phenomenon of spin-down of a
forming neutron star due to neutrino emission during the core
collapse. The spin-down mechanism consists in the fact that
neutrinos which escape from the star are produced in regions
around the neutrinosphere, where the transverse velocity of matter
is relatively lower, and as they propagate to outer regions, some
of them will collide with matter moving with larger transverse
velocities, thus absorbing transverse momentum and causing a spin
down of the medium. In this sense, the trajectory of an outgoing
neutrino should bend as it propagates through the rotating medium,
due to collisions. For this effect to be of any significance, the
mean free path should be less (but not much less) than the star
radius. Otherwise, if the mean free path is much larger, there
will be too few collisions, while if it is much shorter, the
difference in transverse velocities from the emission to the
collision points will be too small.

As cited above, spin-down due to neutrino emission was already
proposed in the past. However, with the knowledge available today
we can include more details in the treatment, namely (i) take
proper account of the weak interactions in the collisions, (ii)
use a phenomenological matter composition and density profile in
the star instead of a uniform medium, (iii) take into account the
opacity of inner parts of the star where neutrinos thermalize, so
for the spin-down effect neutrinos are emitted only from a {\sl
neutrinosphere} instead of from the center of the star, and (iv)
use thermal spectra for each neutrino species instead of a single
monochromatic emission.

We organize the calculation starting from the neutrino production
at the neutrinosphere, followed by the individual neutrino
collisions with the rotating medium further outside, and finally,
by adding these collisions all around the star, to get the result.
At each step we state the model of the situation and the
approximations used.

Neutrinos are produced everywhere around the star, but at high
temperatures (energies) they suffer much scattering and
absorption. Therefore, those which manage to escape are not
produced at the star center but at the neutrinosphere, or surface
of last scattering~\cite{Mez04}. The definition of the
neutrinosphere is statistical and depends on the medium density as
well as the neutrino species and energy. Here we will simply
define it as a spherical shell of radius $R_{ns}$, different for
every neutrino species and energy, given by one mean free path
less than the star radius,
\begin{equation}\label{NSPHERE}
  R_{ns} = R- \frac{1}{\sigma_{\nu} n},
\end{equation}
where $R$ is the star radius and $\sigma_{\nu}$ is the
(energy-dependent) total cross section for neutrino scattering in
the star medium with nucleon number density $n$. Since $n$ depends
on the position, Eq.~\eqref{NSPHERE} is really an equation for
$R_{ns}$ that we solve in each case. We consider a PNS medium
where a fraction $Y_n \sim 0.9$ of the nucleons are neutrons and a
fraction $Y_p \sim 0.1$ are protons~\cite{Lattimer91}.

We neglect scattering with electrons, because they are in equal
number to protons but their cross section is an order of magnitude
smaller. We also neglect Pauli blocking or nucleon correlations.
These effects should be more important at later stages of the star
evolution, when neutrino energies are lower; however, the
spin-down caused by neutrinos is negligible then. In any case,
correlations tend to reduce the cross sections~\cite{Reddy},
pointing further into the direction that previous calculations of
the PNS spin-down by neutrinos were overestimated.

The total cross sections (and thus neutrinospheres) differ for
three species of neutrinos: electron neutrinos $\nu_e$, electron
antineutrinos $\bar\nu_e$, and all other $\nu_x$. The
neutrino-nucleon cross sections, for neutrinos with energies
$1-100$ MeV, are approximately (see pp.~160--167 in
Ref.~\cite{GiuKim07p511})
\begin{align}\label{CROSS_0}
  \sigma (\nu_e n)_\mathrm{inelastic} = &
  \sigma (\bar\nu_e p)_\mathrm{inelastic}
  \notag
  \\
  = & 9.1\times 10^{-42}
  \left(
    \frac{E_\nu}{10\thinspace\text{MeV}}
  \right)^2
  \thinspace\text{cm}^2,
  \notag
  \\
  \sigma (\nu n)_\mathrm{elastic} = &
  2.6\times 10^{-42}
  \left(
    \frac{E_\nu}{10\thinspace\text{MeV}}
  \right)^2
  \thinspace\text{cm}^2,
  \notag
  \\
  & (\text{all $\nu$ species}),
  \notag
  \\
  \sigma (\nu p)_\mathrm{elastic} = &
  2.1\times 10^{-42}
  \left(
    \frac{E_\nu}{10\thinspace\text{MeV}}
  \right)^2
  \thinspace\text{cm}^2,
  \notag
  \\
  & (\text{all $\nu$ species}).
\end{align}
The total cross sections in the star medium for the species in
question are then
\begin{eqnarray}
  \sigma_{\nu_e} &=& Y_n\
  \left[
    \sigma(\nu_e n)_\mathrm{inel.} +\sigma(\nu n)_\mathrm{el.}
  \right]
  + Y_p\
  \sigma(\nu p)_\mathrm{el.},
  \nonumber
  \\
  \sigma_{\bar \nu_e} &=& Y_p\
  \left[
    \sigma(\bar\nu_e p)_\mathrm{inel.} +\sigma(\nu p)_\mathrm{el.}
  \right]
  + Y_n\
  \sigma(\nu n)_\mathrm{el.},
  \nonumber
  \\
  \sigma_{\nu_x} &=& Y_n\
  \sigma(\nu n)_\mathrm{el.} +
  Y_p \
  \sigma(\nu p)_\mathrm{el.},
  \label{TOTCROSS}
\end{eqnarray}
which result in the hierarchy $\sigma_{\nu_e} > \sigma_{\bar\nu_e}
> \sigma_{\nu_x}$. Consequently, the sizes of the respective
neutrinospheres follow the same order and, since the star
temperature is higher further inside, the average energies also
follow a hierarchy~\cite{Tot98}:
\begin{gather}\label{energies}
  \langle E_{\nu_e}\rangle  \sim 10\thinspace\text{MeV},
  \quad
  \langle E_{\bar\nu_e}\rangle  \sim 15\thinspace\text{MeV},
  \notag
  \\
  \langle E_{\nu_x}\rangle  \sim 20\thinspace\text{MeV}.
\end{gather}
Concerning the neutrino spectra, we take two alternative
approaches. In our first, simpler approach, we consider purely
monoenergetic neutrinos for each species, as given in
Eq.~\eqref{energies}. Each species $\nu_i$ has then a definite
neutrinosphere. In our second, more refined approach, we consider
thermal (Fermi-Dirac) energy distributions for each neutrino
species~\cite{Kei03}, with temperatures according to the energy
averages given in Eq.~\eqref{energies}. In this case, the
neutrinospheres are continuously distributed.

We also checked for the possibility that neutrinos could
experience flavor oscillations in matter while propagating from
the neutrinosphere towards the star surface, but found this effect
to be irrelevant for the spin-down. Neutrino resonant conversion
at energies near $10\thinspace\text{MeV}$  is important in the
expanding envelope~\cite{Tomas} where the matter density is
$(10^{26}-10^{27})\thinspace\text{cm}^{-3}$, but in our region of
interest of the star densities are about
$10^{35}\thinspace\text{cm}^{-3}$ (see Ref.~\cite{GiuKim07p511}),
where flavor oscillations are suppressed. Oscillations can happen,
but in a rather thin layer close to the neutron star surface, and
so their effect on the spin-down is negligible.

Another group of flavor changing processes, which can influence
the angular momentum transfer by neutrinos, is the neutrino flavor
conversion due to the $\nu-\nu$ scattering~\cite{NotRaf88,Dua06}.
This effect was shown to significantly change the initial flavor
content of neutrinos in a dense neutrino flux, corresponding to
the neutrino luminosity $\mathcal{L}_\nu >
10^{51}\thinspace\text{erg/s}$. It was however found in
Ref.~\cite{Dua06} that the significant transition probability due
to $\nu-\nu$ collisions is archived at the distances
$(70-80)\thinspace\text{km}$ from the neutrinosphere surface, i.e.
it happens in the envelope of a star. On the contrary, we study
the spin-down of PNS due to the collisions with background matter
in the core of PNS outside the neutrinosphere at the distances
$<20\thinspace\text{km}$ from the star center (see
Fig.~\ref{amtfig} below).

Note that background matter can also influence the collective
neutrino flavor transformations. In order to have some resonance
effects in these flavor changing processes, the neutrino density
$n_\nu (r) = \mathcal{L}_\nu / 4 \pi r^2 \langle E_\nu \rangle$,
where $\langle E_\nu \rangle \sim 10\thinspace\text{MeV}$ is the
typical neutrino energy and $\mathcal{L}_\nu \sim
10^{52}\thinspace\text{erg/s}$, should be comparable with the
electrons number density $n_e$. In our calculations we suggest
that $n_e \approx 0.1 n_n \sim 10^{34}\thinspace\text{cm}^{-3}$
inside the core of PNS at $r<(15-20)\thinspace\text{km}$. The
number density of neutrinos at the same distance is $n_\nu \sim
10^{32}\thinspace\text{cm}^{-3}$, which is 2 orders of magnitude
less than the electron density. It means that the background
matter is unlikely to generate any resonance effects in our case.

\begin{figure*}
  \centering
  \includegraphics[scale=1]{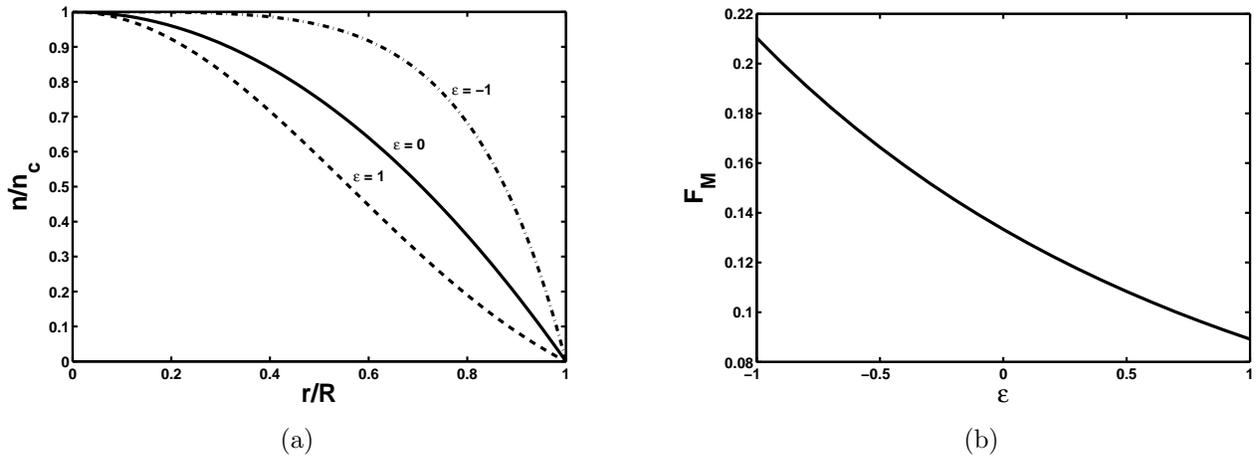}
  \caption{\label{densamt}
  (a) The radial dependence of the density for various values of
  the parameter $\epsilon$. The solid line corresponds to the
  Tolman VII model ($\epsilon = 0$), the dashed line is built for
  $\epsilon = 1$, and the dash-dotted for $\epsilon = -1$.
  (b) The function $F_M(\epsilon)$, given in Eq.~\eqref{totalmass},
  which defines the star mass dependence on the density profile parameter
  $\epsilon$.}
\end{figure*}

Now we should address the density profile of the star. The actual
density profile strongly depends on the equation of state of the
nuclear matter, so these density distributions are generally not
well known~\cite{Neg06}. Results of numerical simulations for the
density profiles~\cite{LatPra01} can be approximated by the
following expression:
\begin{equation}\label{DENSITY}
  n(r)=n_c
  \left(
    1-\frac{r^2}{R^2}
  \right)
  \exp
  \left(
    -\epsilon
    \frac{r^2}{R^2}
  \right),
\end{equation}
where $n_c$ is the central density and $\epsilon$ is a
phenomenological parameter. The $\epsilon=0$ case corresponds to
the well known Tolman VII model~\cite{Tol39}. An analogous
expression for the density profile was used in the study of
nonradial oscillations of a neutron star~\cite{Pod96}. In
Fig.~\ref{densamt}(a) we show the behavior of the density for
various values of the parameter $\epsilon$. In order not to
obscure the sensitivity of our results to different parts of the
calculation, yet  with the risk of not being realistic in specific
cases, we will use these analytical density profiles instead of
full numerical profiles. As shown in Fig.~\ref{densamt}(a),
negative $\epsilon$ implies a more flat profile with a fast drop
at the surface, while positive values imply a gradual decrease of
the density from the center to the surface.

From Eqs.~\eqref{NSPHERE},~\eqref{TOTCROSS} and~\eqref{DENSITY} we
can get the radii of the neutrinospheres as functions of the
parameter $\epsilon$, for a given neutrino energy. In our analysis
we use the values $n_c = (4.7-6.6) \times
10^{38}\thinspace\mathrm{cm}^{-3}$ and $R =
15\thinspace\mathrm{km}$ as well as $n_c = (2.0-2.7) \times
10^{38}\thinspace\mathrm{cm}^{-3}$ and $R =
20\thinspace\mathrm{km}$ for the central density and the star
radius. The total mass of the star calculated on the basis of
Eq.~\eqref{DENSITY} has the form,
\begin{align}\label{totalmass}
  & M \approx
  M_\odot
  \left(
    \frac{n_c}{10^{38}\thinspace\text{cm}^{-3}}
  \right)
  \left(
    \frac{R}{10\thinspace\text{km}}
  \right)^3 F_M(\epsilon),
  \notag
  \\
  & F_M(\epsilon) =
  \int_0^1 \mathrm{d}x x^2 (1-x^2)e^{-\epsilon x^2}.
\end{align}
For the chosen densities and radii Eq.~\eqref{totalmass} gives one
the mass of PNS in the range $(1.4-2.0) M_\odot$, depending on the
parameter $\epsilon$. Although the mass of a neutron star strongly
depends on the equation of state of the neutron star matter, it is
unlikely bigger than $2 M_\odot$~\cite{Hae07}. In
Fig.~\ref{densamt}(b) we present the $\epsilon$ dependence of the
function $F_M(\epsilon)$, which is proportional to the PNS mass.

Now, let us address the issue of emission directions. From the
neutrinosphere, neutrinos are emitted outwards in all directions.
However, as a simplified model for the spin-down effect, we
approximate the emission as purely radial, where ``radial'' is
meant in the frame comoving with the local matter at the neutrino
production point. In reality, of course, neutrinos are produced in
all directions and rigorously one should consider the whole
hemisphere of outgoing directions at every point. However, those
particles emitted more towards the tangential velocity will
transfer less momentum to the star under subsequent scattering,
while those emitted against the tangential velocity should
transfer more momentum. In order to test the validity of the
radial emission model, we performed the calculations in the case
of neutrinos emitted in all directions within the equatorial plane
and checked that indeed collisions with purely radially emitted
neutrinos represent the full spin-down effect up to a geometrical
factor near unity. Since there is no need for precision higher
than a few tens of a percent, in what follows we will present the
estimates within the model of radially produced neutrinos.

The process is then calculated by considering that neutrinos are
emitted from their corresponding neutrinosphere and subsequently
collide with the star medium on their way out, thus taking away
part of the angular momentum.

Now let us describe the subsequent elastic collision of the
outgoing neutrino with the star matter, which is where the spin
down effect takes place. The effective weak interaction for
elastic neutrino scattering with a background fermion $f$ has the
form~\cite{Oku90},
\begin{align}\label{matrellscat}
  M = & \frac{G_\mathrm{F}}{\sqrt{2}}
  \bar{f}(p_2)
  [g_L \gamma_\mu (1-\gamma^5) + g_R \gamma_\mu (1+\gamma^5)]
  f(p_1)
  \notag
  \\
  & \times
  \bar{\nu}(k_2) \gamma^\mu (1-\gamma^5) \nu(k_1),
\end{align}
where $\nu(k_{1,2})$ are the initial and final neutrino spinors
with momenta $k_{1,2}=(\omega_{1,2},\mathbf{k}_{1,2})$, and
$f(p_{1,2})$ are the respective spinors of the fermions in the
medium, with momenta $p_{1,2}=(E_{1,2},\mathbf{p}_{1,2})$
respectively. The coefficients $g_{L,R}$ depend on the neutrino
scattering channel and are listed in Table~\ref{gLgRtab}.
\begin{table}
\begin{center}
  \caption{The values of the coefficients $g_{L,R}$ in
  Eq.~\eqref{matrellscat} for various channels of neutrino elastic
  scattering. $\nu_x$ stands for $x\neq e$ while $\nu$ stands for all
  lepton flavors.\label{gLgRtab}}
  \begin{ruledtabular}
  \begin{tabular}{ccrr}
    No. & Reactions & $g_L$ & $g_R$ \\
    \hline
    1 & $\nu_e e \to \nu_e e$ & 0.73 & 0.23 \\
    2 & $\nu_x e \to \nu_x e$ & -0.27 & 0.23 \\
    3 & $\bar{\nu}_e e \to \bar{\nu}_e e$ & 0.23 & 0.73 \\
    4 & $\bar{\nu}_x e \to \bar{\nu}_x e$ & 0.23 & -0.27 \\
    5 & $\nu p \to \nu p$ & 0.27 & -0.23 \\
    6 & $\nu n \to \nu n$ & -0.50 & 0.00 \\
    7 & $\bar{\nu} p \to \bar{\nu} p$ & -0.23 & 0.27 \\
    8 & $\bar{\nu} n \to \bar{\nu} n$ & 0.00 & -0.50 \\
  \end{tabular}
  \end{ruledtabular}
\end{center}
\end{table}

From the effective interaction of Eq.~\eqref{matrellscat}, we can
build the cross section,
\begin{align}\label{dsigmak2p2}
  \mathrm{d}\sigma (\nu f\to \nu f) = & \frac{1}{64 \pi^2}
  \delta^4({p}_2+{k}_2-{p}_1-{k}_1)
  \notag
  \\
  & \times
  \frac{|M|^2}{(k_1\cdot p_1)\ \omega_2 E_2}
  \mathrm{d}^3\mathbf{k}_2 \mathrm{d}^3\mathbf{p}_2,
\end{align}
where the matrix element squared for unpolarized scattering
derived from Eq.~\eqref{matrellscat} is
\begin{align}\label{M2}
  |M|^2 = & 128 G_\mathrm{F}^2[\, g_L^2(p_1\cdot  k_1)^2 + g_R^2(p_1\cdot  k_2)^2
  \notag
  \\
  & -
  g_L\, g_R\, m_f^2 (k_1\cdot  k_2)\, ],
\end{align}
with $m_f$ being the mass of the fermion $f$.
For radially emitted neutrinos, and taking this radial direction
at each collision point as our reference, the incoming fermion
from the medium will be perpendicular and the outgoing neutrino
will have an arbitrary direction parametrized by the relative
polar and azimuthal angles $\theta_2$ and $\phi_2$, such that
\begin{align}\label{3momgeom}
  \mathbf{k}_1 = & \omega_1 (0,0,1),
  \quad
  \mathbf{p}_1  = E_1 v_f (0,1,0),
  \notag
  \\
  \mathbf{k}_2 = & \omega_2
  (\sin\theta_2\cos\phi_2, \sin\theta_2\sin\phi_2, \cos\theta_2).
\end{align}
Here $v_f$ is the tangential velocity of the star at the
interaction point. We are assuming that the thermal velocities of
the fermions in the medium average out concerning this effect, and
the only effective velocity is the average drift of the medium due
to rotation. With these conventions, the dot products of interest
are
\begin{align}\label{4momgeom}
  (p_1\cdot k_1) = & E_1\omega_1,
  \quad
  (k_1\cdot k_2) = \omega_1 \omega_2 (1-\cos\theta_2),
  \notag
  \\
  (p_1\cdot k_2) = & E_1 \omega_2 (1-v_f \sin\theta_2\sin\phi_2).
\end{align}
We can integrate Eq.~\eqref{dsigmak2p2} over $\mathbf{p}_2$ to get
the differential cross section,
\begin{equation}\label{dsigmanu}
  \frac{\mathrm{d}\sigma}{\mathrm{d}\Omega_2} =
  \frac{|M|^2}{64 \pi^2}
  \frac{\omega_2 ^2}{\omega_1 ^2 E_1 ^2},
\end{equation}
where the energy of the outgoing neutrino is given, by
energy-momentum conservation, in terms of the scattering angles
\begin{align}\label{paramscatt}
  \omega_2 = &
  \frac{E_1\omega_1}{E_1(1-v_f \sin\theta_2\sin\phi_2)+\omega_1(1-\cos\theta_2)}.
\end{align}

Now we can calculate the rate of transverse momentum transferred
to the star at that collision point. Let $J(r)$ be the (radial)
neutrino flux incoming to a given collision point; then $J(r)
(\mathrm{d}\sigma/\mathrm{d}\Omega_2)\, \mathrm{d}\Omega_2$ is the
rate of outgoing neutrinos within $\mathrm{d}\Omega_2$. Since the
$\mathbf{e}_\phi$ component of momentum of each of these outgoing
neutrinos is $\omega_2\sin\theta_2 \sin\phi_2$, then the total
$\mathbf{e}_\phi$ momentum per unit time transferred to the star
due to the outgoing neutrinos ejected from a given collision point
is
\begin{equation}\label{kphi}
  \langle \dot{k}_\phi \rangle =
  \int \omega_2 \sin\theta_2 \sin\phi_2 J(r)
  \frac{\mathrm{d}\sigma}{\mathrm{d}\Omega_2} \ \mathrm{d}\Omega_2  ,
\end{equation}
where $\mathrm{d}\Omega_2 = \sin\theta_2 \mathrm{d}\theta_2
\mathrm{d}\phi_2$. The flux $J(r)$ at the collision point is
related to the neutrino flux at the surface of the star, $J_0$, by
the relation $J(r)=J_0\times(R/r)^2$, where $R$ is the star
radius. Using Eq.~\eqref{kphi} we can now calculate the total rate
of angular momentum transferred to the star as a whole by just
summing over all collision points that lie outside the
corresponding neutrinosphere,
\begin{equation}\label{Lz}
  \dot{L}_z =
  \int_{r>R_{ns}} \langle \dot{k}_\phi \rangle r \sin\vartheta\  n_f(\mathbf{r}) \
  \mathrm{d}^3\mathbf{r},
\end{equation}
where $n_f(\mathbf{r})$ is the local number density of target
fermions in the medium, and $\vartheta$ is the polar angle
(colatitude) of  the star at the collision point. The full result
for $\dot{L}_z$ is the sum of this type of calculation, repeated
for each neutrino species.

The rate of angular momentum transfer to the star as given in
Eq.~\eqref{Lz} requires the computation of a four-fold integral
which has to be evaluated numerically.
Nevertheless, within the following approximations one can
calculate an analytical expression for $\dot{L}_z$.
First, assuming a star radius  $R= 20\thinspace\mathrm{km}$ and
angular velocity $\Omega = 10^3\thinspace\mathrm{s}^{-1}$, its
equatorial linear velocity (in units of $c$) is $\sim 0.1$. Thus
we can treat $v_f$ in Eq.~\eqref{paramscatt} as a small parameter.
In addition, typical energies for the emitted neutrinos  are
$\omega_1 \sim 10\thinspace\mathrm{MeV}$, while the incoming
fermion in the medium, which is nonrelativistic, has an energy
$E_1$  near the nucleon mass. Therefore the ratio $\omega_1/E_1$
is also small. Consequently, we use the approximation,
\begin{align}\label{parscat}
  \omega_2 \approx & \omega_1(1+\xi),
  \\
  \intertext{where}
  \notag
  \xi = &
  v_f\sin\theta_2\sin\phi_2-\frac{\omega_1}{E_1}(1-\cos\theta_2)
  \ll 1.
\end{align}
For the matter velocity at the collision point we use
\begin{equation}\label{vf}
  v_f=v_0 \sin\vartheta \ \frac{r-R_{ns}}{R},
\end{equation}
where $v_0$ is the equatorial velocity of the neutron star.
Here $v_f$ corresponds to the \emph{relative} transverse velocity
of the medium at the collision point (radius $r$) with respect to
the velocity of the medium at the neutrino emission point (radius
$R_{ns}$).

Using Eq.~\eqref{M2} and Table~\ref{gLgRtab}, we can determine the
square of the matrix element, for example for the specific
reaction $\nu_e n \to \nu_e n$, which is $|M|^2=32 G_\mathrm{F}^2
E_1 ^2 \omega_1 ^2$. Consequently, using Eqs.~\eqref{Lz}
and~\eqref{parscat} one gets the following approximation for the
rate $\dot{L}_z$ due to this type of collision:
\begin{align}\label{Lzapp}
  \dot{L}_z \approx &
  2 G_\mathrm{F}^2 \omega_1^3 R J_0 v_0
  \int_{R_{ns}}^R r(r-R_{ns}) \mathrm{d}r
  \notag
  \\
  & \times
  \int_0^\pi n_n(r,\vartheta)\sin^3\vartheta\, \mathrm{d}\vartheta,
\end{align}
where $n_n(r,\vartheta)$ is the neutron density at the
corresponding collision point.

It is convenient to present the final result as the ratio between
the rate of angular momentum loss and the initial angular momentum
of the star $L_0 = \mathcal{I}\, \Omega$, where its moment of
inertia is given by
\begin{equation}\label{L0}
  \mathcal{I} = \frac{8\pi}{3} m_n
  \int_0^R \mathrm{d} r\,  r^4\, n(r),
\end{equation}
and where $m_n$ is the neutron mass. Based on Eqs.~\eqref{DENSITY}
and~\eqref{Lzapp} we find for this ratio:
\begin{align}\label{dLzLNS}
  \frac{\dot{L}_z}{L_0} = &
  \frac{G_\mathrm{F}^2 \mathcal{L}_\nu E_\nu^2}{4 \pi^2 R^2 m_n}
  \frac{F_1(\epsilon)}{F_0(\epsilon)}
  \notag
  \\
  & \approx
  1.0\thinspace\mathrm{s}^{-1} \times
  \left(
    \frac{E_\nu}{10\thinspace\mathrm{MeV}}
  \right)^2
  \left(
    \frac{\mathcal{L}_\nu}{10^{52}\thinspace\mathrm{erg}/\mathrm{s}}
  \right)
  \notag
  \\
  & \times
  \left(
    \frac{R}{10\thinspace\mathrm{km}}
  \right)^{-2}
  \frac{F_1(\epsilon)}{F_0(\epsilon)},
\end{align}
where $\mathcal{L}_\nu$ is the neutrino luminosity and where we
have defined the following integrals:
\begin{align}\label{F0F1}
  F_0(\epsilon) = & \int_0^1 \mathrm{d}x\, x^4 (1-x^2) e^{-\epsilon\, x^2},
  \notag
  \\
  F_1 (\epsilon) = & \int_{x_0}^1 \mathrm{d}x\, x (x-x_0) (1-x^2) e^{-\epsilon\, x^2},
\end{align}
with $x_0 \equiv R_{ns}/R$.

We have just considered the reaction $\nu_e n \to \nu_e n$. The
contributions of all other neutrino species, including collisions
with protons, are treated in a similar way and added to the
result. The average neutrino luminosity just after the
neutronization stage~\cite{Tot98} is $\mathcal{L}_\nu \sim
10^{52}\thinspace\mathrm{erg}/\mathrm{s}$.
This large neutrino luminosity lasts for a few seconds, mainly
during the Kelvin-Helmholtz stage, with all neutrino types,
$\nu_{e,\mu,\tau}$ and $\bar{\nu}_{e,\mu,\tau}$, having almost
equal luminosities, and therefore giving similar contributions to
Eq.~\eqref{dLzLNS}.

A refinement of this neutrino emission model considers the three
neutrino species not monoenergetic but with thermal distributions,
each with distinct temperatures and chemical potentials. Therefore
we can choose the Fermi-Dirac distribution in the
form~\cite{Kei03}
\begin{align}\label{Edistr}
  \frac{\mathrm{d}N}{\mathrm{d}E_\nu} = &
  \frac{\mathcal{L}_\nu}{F(\eta_\nu) T_\nu^4}
  \frac{E_\nu^2}{\exp(E_\nu/T_\nu - \eta_\nu)+1},
  \notag
  \\
  F(\eta_\nu) = & \int_0^\infty \mathrm{d}x \frac{x^3}{e^{x-\eta_\nu}+1},
\end{align}
where we have the relation $\langle E_\nu \rangle / T_\nu \approx
3.1514 + 0.1250 \eta_\nu + 0.0429 \eta_\nu^2 + \dotsb$ between the
temperature of the neutrino gas $T_\nu$ and the mean neutrino
energy $\langle E_\nu \rangle$ defined in Eq.~\eqref{energies}.
The typical values of the chemical potentials are (see
Ref.~\cite{Kei03}) $\eta_{\nu_e} \approx 2$, $\eta_{\bar{\nu}_e}
\approx 3$, and $\eta_{\nu_x} \approx 1$.

On the basis of Eqs.~\eqref{dLzLNS}-\eqref{Edistr} we get the
averaged ratio as
\begin{align}\label{dLzLNSthermal}
  \left\langle
    \frac{\dot{L}_z}{L_0}
  \right\rangle \approx &
  1.0\thinspace\mathrm{s}^{-1}\times
  \left(
    \frac{T_\nu}{10\thinspace\mathrm{MeV}}
  \right)^{-4}
  \notag
  \\
  & \times
  \left(
    \frac{\mathcal{L}_\nu}{10^{52}\thinspace\mathrm{erg}/\mathrm{s}}
  \right)
  \left(
    \frac{R}{10\thinspace\mathrm{km}}
  \right)^{-2}
  \notag
  \\
  & \times
  \frac{1}{F(\eta_\nu)}
  \int_0^\infty
  \frac{y^5\mathrm{d}y}{e^{y-\eta_\nu}+1}
  \frac{F_1(\epsilon,y)}{F_0(\epsilon)},
\end{align}
where we are including the energy dependence of the function
$F_1(\epsilon)$ in Eq.~\eqref{F0F1} in the form of the
dimensionless parameter $y = (E_\nu/10\thinspace\text{MeV})$,
since the size of the neutrinosphere depends on the neutrino
energy [see Eqs.~\eqref{NSPHERE} and~\eqref{CROSS_0}].

In Fig.~\ref{amtfig}
\begin{figure*}
  \centering
  \includegraphics[scale=.95]{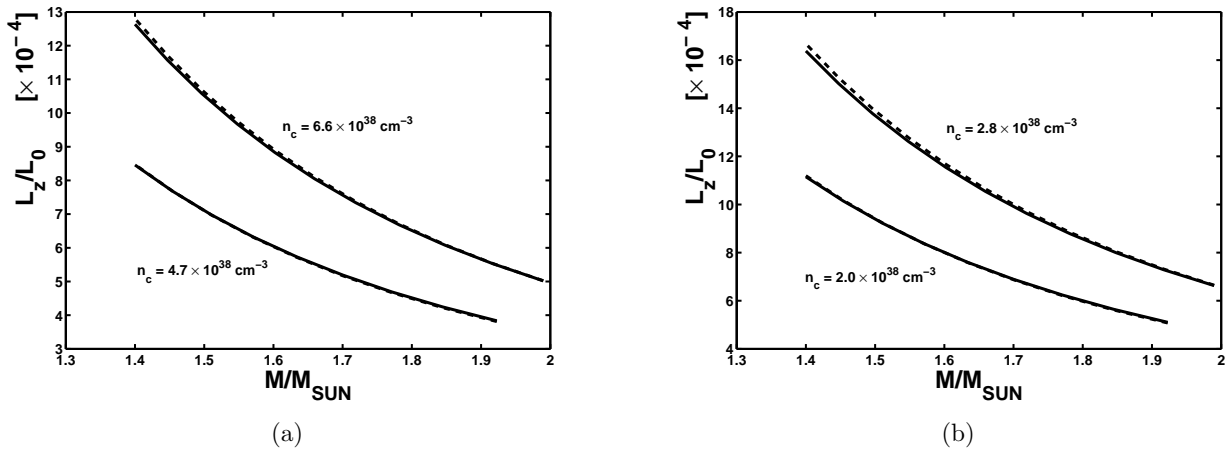}
  \caption{\label{amtfig}
  The fraction of the total angular momentum carried away by
  all neutrinos types versus the PNS mass, which depends on the parameter
  $\epsilon$ according to Eq.~\eqref{totalmass}, for
  various central densities. The dashed lines correspond to the simple
  model~\eqref{dLzLNS} with monoenergetic neutrinos with
  energies given in Eq.~\eqref{energies}. The solid lines correspond to
  the refined model~\eqref{dLzLNSthermal} with
  thermally distributed neutrinos with average energies
  (temperatures) given in Eq.~\eqref{energies}.
  (a) The fraction for a star with $R = 15\thinspace\text{km}$.
  The upper curves correspond to
  $n_c = 6.6 \times 10^{38}\thinspace\text{cm}^{-3}$ and the lower
  ones to $n_c = 4.7 \times 10^{38}\thinspace\text{cm}^{-3}$.
  (b) The fraction for a star with $R = 20\thinspace\text{km}$.
  The upper curves correspond to
  $n_c = 2.7 \times 10^{38}\thinspace\text{cm}^{-3}$ and the lower
  ones to $n_c = 2.0 \times 10^{38}\thinspace\text{cm}^{-3}$.}
\end{figure*}
we present the total angular momentum  carried away by all
neutrino species, $L_z = \sum_i \dot{L}_z(\nu_i) \Delta t$, where
$\Delta t = 10\thinspace\mathrm{s}$ , in units of the initial
angular momentum $L_0$, versus the mass of PNS. Here $\Delta t$ is
the time interval during which the luminosity stays at the high
value $\mathcal{L}_\nu \approx
10^{52}\thinspace\mathrm{erg}/\mathrm{s}$. Note that the PNS mass
variation on this plot results from its epsilon dependence [see
Eq.~\eqref{totalmass} and Fig.~\ref{densamt}(b)].

As one can see from the figures, a larger fraction of the angular
momentum can be carried away by neutrinos for a star with bigger
radius. We can also notice that with the enhancement of the
central density the effect increases. On one hand, the
neutrinosphere radius grows with the enhancement of the central
density~\eqref{NSPHERE}. Thus neutrinos have less opportunity to
collide with rotating matter and the relative velocities from
production to collision are also smaller. On the other hand, PNS
with equal masses and different central densities correspond to a
different $\epsilon$ parameter in Eq.~\eqref{totalmass}. For
example, $n_c = 4.7 \times 10^{38}\thinspace\text{cm}^{-3}$
corresponds to $\epsilon = 1$ and $n_c = 6.6 \times
10^{38}\thinspace\text{cm}^{-3}$ to $\epsilon = 2$ for $M = 1.4
M_\odot$. The bigger value of $\epsilon$ implies a more steep
density profile near the central region and, hence, a bigger
concentration of PNS mass there. It means that effectively a
neutrinosphere should decrease at higher $\epsilon$. The results
of numerical simulations presented in Fig.~\ref{amtfig} shows that
the latter effect is more significant.

One can also see in Fig.~\ref{amtfig} that the difference between
the models of monoenergetic and thermally distributed neutrinos is
less than several percent. Thus, a monoenergetic neutrino model is
still a good approximation to describe the spin-down.

Including general relativity corrections to the moment of inertia
of the neutron star, the values of $\dot{L}_z/L_0$ become slightly
smaller than those we obtained. For example, if we consider a
neutron star with $M = 2\,M_\odot$, the reduction factors within
the Tolman VII model ($\epsilon = 0$) will be $\approx 0.76$ for
$R = 15\thinspace\mathrm{km}$ and $\approx 0.82$ for $R =
20\thinspace\mathrm{km}$~\cite{LatPra01}. As a result, we find
that, by the mechanism we have considered, and within the Tolman
VII model as benchmark ($\epsilon = 0$) for a
$R=15\thinspace\mathrm{km}$ star, neutrinos could carry away up to
$2.4 \times 10^{-4}$ of the initial angular momentum of the PNS
[compare this value to the corresponding result shown in
Fig.~\ref{densamt}(a), which does not include the general
relavivity corrections].

For other density profiles, i.e., corresponding to $M = 1.4
M_\odot$ ($\epsilon=2$), neutrinos can carry away $1.7 \times
10^{-3}$ of the initial angular momentum for a star with $R =
20\thinspace\mathrm{km}$ and $n_c = 2.8 \times
10^{38}\thinspace\text{cm}^{-3}$. This result is significantly
less than predicted in the previous works~\cite{Mik77,Eps78a}.

What follows at later stages of the neutron star evolution is that
the neutrinospheres shrink as the star cools down, because  the
mean free path grows beyond the star radius. This is mainly due to
the Pauli blocking reduction of the neutrino scattering cross
section. At the same time, the neutrino flux drops down
dramatically from its initial values. The expression in
Eq.~\eqref{dLzLNS} is then valid only at the initial stages, which
is the only time when the spin-down due to neutrino emission can
be of any significance.


\section{Summary and Conclusions\label{CONCL}}

We have studied the spin-down of a forming neutron star due to
neutrino emission. The spin-down by neutrino emission is
significant only for the first few seconds of the neutron star
evolution, when the size of the neutrinosphere is less than (but
comparable to) the radius of the star and the neutrino flux is
still large. These conditions allow for neutrinos to have a large
enough collision rate. At later stages the star becomes almost
transparent for neutrinos and the neutrino flux is too small for
this effect to be of any significance, so other spin-down
mechanisms take over.

We model the phenomenon by considering neutrinos of type $\nu_e$,
$\bar\nu_e$, and $\nu_x$ (where $x$ stands for all other), which
have different interaction with the medium, each produced at their
corresponding neutrinosphere and moving radially outwards,
subsequently colliding with the star matter in the outer shells,
where the transverse velocity of the medium is larger than at the
production point, thus causing a slowing down of the star.

An arguable part of the model is the consideration of purely
radially moving neutrinos at the production points, but we have
checked analytically in some simplified cases that the emission in
all directions causes an average effect not much different than
the purely radial case.

Another simplification is that Pauli blocking and other nucleon
correlations are not taken into account, presuming that they are
more important at later stages, when neutrino flux and energies
are lower and this spin-down mechanism is negligible.

Another important part of the model concerns the density profile,
where we use a phenomenological analytical expression instead of
more realistic numerical profiles, in order to study the
sensitivity to it and have a better comparison to previous
estimates to the spin-down phenomenon. We find that the density
distribution, $n(r)$, significantly affects the results for the
rate of angular momentum loss (see Fig.~\ref{densamt}). Very few
analytical formulas for $n(r)$ models are known, and the majority
of the density profiles are available from numerical simulations.
We have chosen an exponential density profile~\eqref{DENSITY}
which depends on a phenomenological parameter $\epsilon$ to fit
the results of the numerical simulations~\cite{LatPra01}. For
$\epsilon = 0$ this formulation reproduces the well known Tolman
VII model. Although this model cannot capture the fine details of
individual cases, it reproduces the main features of realistic
stars. However, it is unable to explain without fine tuning the
enormous range of matter densities, from $\sim
10^6\thinspace\mathrm{g/cm}^3$ in the crust to $>
10^{14}\thinspace\mathrm{g/cm}^3$ in the center. Therefore we have
introduced an additional exponential factor in the model for
$n(r)$.

As shown in Fig.~\ref{densamt}(a), for positive $\epsilon$ the
density goes gradually down, while for negative $\epsilon$ it
stays high almost all over the star, falling sharply at the
surface. Consistently, Fig.~\ref{amtfig} shows that the spin-down
effect is bigger for PNS with smaller mass at the fixed central
density. Taking into account Fig.~\ref{densamt}(b) we obtain that
spin-down should increase with the enhancement of $\epsilon$
because a higher and flatter density implies a larger
neutrinosphere radius, thus decreasing the relative velocity of
the star at the collision point with respect to the point of
production. On the contrary, if neutron stars would feature a more
gradual density descent, the average distance from the
neutrinospheres to the collision points will increase, causing a
stronger spin-down effect.

We have found that, for some density profiles corresponding to
$\epsilon=2$ in the case of a PNS with $R = 20\thinspace\text{km}$
and central density $n_c = 2.8 \times
10^{38}\thinspace\text{cm}^{-3}$, which gives one $M = 1.4
M_\odot$, neutrinos can carry away up to about $1.7 \times
10^{-3}$ of the initial angular momentum of the neutron star,
provided that an average neutrino luminosity near
$\mathcal{L}_\nu=10^{52}\thinspace\mathrm{erg}/\mathrm{s}$ lasts
for about $10\thinspace\mathrm{s}$ during the Kelvin-Helmholtz
stage of the neutron star evolution. The results of our
calculations for other central densities and radii are presented
in Fig.~\ref{amtfig}.

We should recall that previous estimates~\cite{Mik77,Eps78a} gave
larger values of $\dot{L}_z$, and it was even predicted that the
rotation of a neutron star could be stopped by neutrino emission
(or equivalently, a reduction of the angular velocity by more than
an order of magnitude). After using a variable density profile,
our better knowledge of electroweak interactions today, and the
most recent data on the neutrino flux, we find that the results
are clearly not as dramatic.

Let us examine the importance of various factors which were not
accounted for in the previous calculations of the PNS spin
down~\cite{Mik77,Eps78a}. For example, we can discuss the
situation when all neutrinos are monoenergetic, with $E_\nu \sim
10\thinspace\text{MeV}$ and $\mathcal{L}_\nu \sim
10^{52}\thinspace\text{erg/s}$, and are emitted from the PNS
center, whereas the density profile is given by
Eq.~\eqref{DENSITY}. The considered case is equivalent to the
neutrinosphere with zero radius: $x_0 = 0$ in Eq.~\eqref{F0F1}.
Using Eq.~\eqref{dLzLNS} for PNS with $R \sim (15 -
20)\thinspace\text{km}$, we obtain that $\dot{L}_z \Delta t \sim
L_0$, where $\Delta t \sim \text{several seconds}$.
Therefore we obtain that the rotation of PNS can be significantly
reduced by the neutrino emission provided all the particles are
emitted in the center of PNS, which reproduces the results of
Refs.~\cite{Mik77,Eps78a}. It means that the concept of the
neutrinosphere is the most important in our calculations.

It can be also shown that various density distribution profiles
can change the PNS spin-down, however not so dramatically.
Concerning other corrections, the inclusion of Pauli blocking or
nucleon correlations would tend to make the effect even smaller;
different proportions of neutrino species or neutrino oscillations
only induce minor or negligible changes; using monoenergetic or
thermally distributed neutrinos also result in minor differences
in the spin-down effect.

Concerning the experimental observation of this effect,
unfortunately there is quite limited information about the initial
angular velocities of neutron stars. An effort to infer initial
angular velocities of PNS has been made~\cite{SwaWu01}, where it
was revealed that there should be a large uncertainty in the
results.

\begin{acknowledgments}
  This work has been supported by Fondecyt (Chile) Grant No. 1070227 and
  Conicyt (Chile), Programa Bicentenario PSD-91-2006. We also thank
  L.~B.~Leinson, J.~Maalampi, G.~G.~Raffelt, A.~Reisenegger,
  I.~Schmidt and A.~I.~Studenikin for helpful discussions.
  One of the authors (MD) is grateful to
  Deutscher Akademischer Austausch Dienst for a grant.
  The comments of the referee are also appreciated.
\end{acknowledgments}

\end{document}